\documentclass[preprint,aps,floats,showpacs,nofootinbib]{revtex4}
\usepackage{graphics}
\begin{document}
\title{Enhancement of the gluonic couplings of the Standard Model Higgs Boson through
mixing}
\preprint{OSU-HEP-01-06}
\author{S. Nandi {\it and\/} T. Torma}
\affiliation{Department of Physics, Oklahoma State
University\\Stillwater OK 74078, U.S.A.}
\date{\today}
\begin{abstract}
The mixing of the Standard Model Higgs boson with a $t\overline t$ scalar bound state
(arising from a new stronger-than-QCD chiral interaction) can enhance the Higgs boson's coupling to
two gluons. It is possible that the two-gluon decay mode of a light Higgs boson may even dominate.
This alters the Higgs boson search strategy at Tevatron Run~2 in a significant way. We present a
simple model in which such a scenario may be realized. The modifications to the Higgs search signal,
as well as further experimental tests of this scenario are also discussed.
\end{abstract}
\maketitle
\section{Introduction}

One of the major objectives of Run~2 at the Tevatron is to search for the Higgs boson in the
mass range of $110\,GeV$ to $180\,GeV$~\cite{TevRep}. Direct searches at LEP2 have shown no clear
evidence of the Higgs boson, and have set a lower limit of $114.1\,GeV$ for the Standard Model
(SM) Higgs boson~\cite{N2}. Global analyses of LEP1 electroweak data have considerably narrowed
down the SM Higgs boson mass~\cite{N3}, and prefer a low mass Higgs boson. At Tevatron energy
($\sqrt s=2\,TeV$), the dominant mechanism of SM Higgs boson production is $gg\to h$ with a cross
section ranging from
$0.9\,pb$ for $m_h=110\,GeV$ to $0.2\,pb$ for $m_h=180\,GeV$. For $m_h<135\,GeV$, $h$
decays dominantly to $b\overline b$ and this mode of production and decay is hopeless due to the
overwhelming QCD background. Form $m_h>135\,GeV$, the $h\to WW^*$ decay mode becomes
dominant, and the detection of the Higgs boson is promising via this mode. The detection via
$\tau^+\tau^-$ decay mode has not been thoroughly studied and may be on the
borderline~\cite{R4}. In the usual scenario, the best way to observe the signal of the SM Higgs boson
for $m_h<135\,GeV$ is the production of $Wh$ via $q\overline q^\prime$ and the subsequent
decay
$h\to b\overline b$, where the background can be controlled using appropriate cuts.
However, if $h\to b\overline b$ is not the dominant decay (as in the scenario we shall
propose), this signal may be significantly lowered, making the detection in this channel very
difficult. On the other hand, if the $ggh$ coupling is significantly enhanced, Higgs boson production
through gluon fusion may be large enough to detect a signal via the $\tau^+\tau^-$ decay mode for
the low mass region ($m_h<135\,GeV$), or the signal in the $WW^*$ decay mode could be large
enough to detect in the higher mass range ($m_h>135\,GeV$). The object of this work is to
explore an alternative scenario in which the $ggh$ coupling is significantly enhanced, and
how that alters the Higgs boson search strategy at Tevatron Run~2 due to the changes of
the relative importance of its various production and decay modes.

In the Minimal SM the Higgs boson has a small coupling to two gluons, generated only at one-loop
level, a fact that makes its discovery difficult on hadron colliders. In extensions to the SM various
Higgs-like scalars are predicted, in some cases with enhanced gluonic
couplings.\footnote{ Composite
Higgs bosons can couple to gluons directly through their quark constituents. The
Higgs mass is typically larger in these models than the range considered in the present
paper. See~\cite{R5}, however, for an example where $m_h\leq330\,GeV$.}
These extensions are pursued in order to
have specific explanations for EW breaking. In this paper we point out that appropriate new physics
can generate an enhancement to this coupling even if mechanism of electroweak breaking is left
intact. The additional interactions present in such models are {\it not required\/} in order to explain
EW breaking, but they cannot be excluded either. Such scenarios must be taken into consideration
because they have important phenomenological consequences.

Such an enhancement arises if the Higgs boson mixes with a $t\overline t$ bound state. Such a state
is not formed in the minimal SM because the top quarks would decay before the strong
interaction could bind it\footnote{The effects of QCD interactions on a $t\overline t$ system
are small. Approximating the QCD effects by a Coulomb potential, we would find
a bound state whose revolution time~\cite{Bigi} is
\(\tau_{r}=\frac{9}{4m_t\alpha_s^2}\sim1\,GeV^{-1}\). Because the top quarks constituting the
bound state have a decay width $\Gamma_t=1.2\,GeV$, the would-be bound state decays before
even one revolution.} (i.e. $\alpha_s(m_t)\sim0.12$ is too weak). However, the right handed top
quark may participate in additional, chiral, new strong interactions. Such interactions are not
phenomenologically excluded if the new gauge symmetry is broken and the new gauge boson
picks up a mass ${\cal O}\left(m_{new}\right)$.\footnote{An interaction in which left handed quarks
participate is excluded by their contribution $Z\to b\overline b$, provided $SU(2)_W$ is respected.
Anomaly cancellation might be arranged through a new fermion sector which may be too heavy to be
presently observed.}
\begin{figure}[tb]
\includegraphics{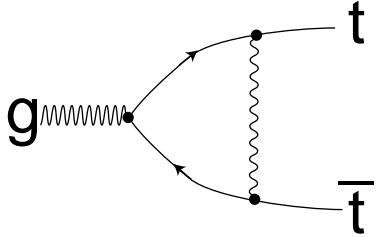}
\caption{\protect\label{fig:corrections}The exchange of a new gauge boson would generate a large
contribution to the $gt\overline t$ coupling unless suppressed by $m_{new}\gg100\,GeV$.}
\end{figure}
We denote by $\sigma$ a hypothetical $t\overline t$ bound state that mixes with the Higgs
boson.  In spectral notation referring to the $t\overline t$ system this state must be
a~$2^{\,3}\!P_0$ state (to ensure even parity and zero spin.) Such a bound state must be held
together by new interactions, whose strength, $\alpha_{new}$, must be much larger than
$\alpha_s\sim 0.12$ to form a bound state at all. They must act only on the third family (otherwise
they would have already been detected in processes such as $q\overline q\to b\overline b$ at the
Tevatron.) The dangerous place is the $gt\overline t$ vertex, where one loop corrections, due to the
exchange of the new gauge bosons might have a large effect, as shown in Fig.~\ref{fig:corrections}. At
the Tevatron, only the $t\overline t$ production cross section will be
affected. This cross section  is known with a $\sim 10\%$ precision, and the effect is
\mbox{${\cal O}\left(\frac{\alpha_{new}}{4\pi}\times\frac{m_t^2}{ m_{new}^2}\right)$}, so we
would prefer a new interaction effective at scales much above~$\sim100\,GeV$.

In this paper our goal is not to build a specific model. Nevertheless we point out that the new
interactions, being chiral, will be very different from a scaled-up copy of QCD. A consequence of this
fact is that the standard way of calculating quarkonium decays, Nonrelativistic QCD~\cite{R6}, is not
applicable here (the size of the NRQCD operators are determined by QCD dynamics.) We'll do
a potential model calculation instead. We use a Yukawa potential to model the $t\overline t$
interaction and argue that the strength of the Yukawa potential will be related to $\alpha_{new}$,
and that the screening length~$m^{-1}$ is related to $m_{new}^{-1}$. These parameters, however,
may not exactly co\"\i ncide, leaving us room to argue that the screening length $m^{-1}$ in the
potential may be somewhat longer than the inverse mass $m_{new}^{-1}$ suppressing visible effects
in top production at the Tevatron.

In Sec.~\ref{sec:mixing} we work out the general formalism of the mixing between a light
Higgs $h$ and the $\sigma$, and show that, with the exception of a small part of the
parameter space, the gluonic branching ratio of the Higgs is determined directly by the
gluonic branching ratio of the $\sigma$ and also by the masses of the $h$ and $\sigma$
(and is {\it independent} of any mixing parameters.) Then we proceed in Sec.~\ref{sec:Yukawa} to
calculate the binding energy and wave functions of the lowest lying states. We find that, in order for
a~$2^{\,3}\!P_0$ state to form and not to have a too wide $\sigma\to b\overline b$ width,
the interaction strength is restricted to be between  \(0.5\leq\alpha_{new}\leq1.0\), while the
screening mass can still exceed $m>100\,GeV$. However, in such a case the binding of the $1S$
states is so deep that their masses become negative. The solution is spin-dependent
potentials. Nuclear physics is witness that strong interactions, even nonchiral, may generate very
strong ${\bf L}\cdot{\bf S}$ couplings. A spin-orbit interaction that binds states with negative values
of ${\bf L}\cdot{\bf S}$ stronger, can lead to a situation when the only bound state that is formed is
$\sigma\equiv2^{\,3}\!P_0$, as we show in a numerical example. We do not think that a one-loop
calculation of the relative strengths of the spin independent and the ${\bf L}\cdot{\bf S}$ terms in
the potential is reliable, so that we handle this ratio as an unknown input parameter.\footnote{We
also note that higher order (in $\alpha_{new}$) effects and/or relativistic corrections may also
induce a mixing between the Higgs boson and the $1^1S_0$ state. At tree level such mixing is absent
and the interaction is repulsive in the singlet and attractive in the triplet channel which does not
allow the formation of a $1^1S_0$ state. This conclusion may be incorrect though when higher
orders are incorporated.} 

In Sec.~\ref{sec:modif} we argue that the presence of such mixing can lead to the
enhancement of final states with two gluons from Higgs decay in lepton colliders. A large
$h\to gg$ branching ratio can lead to a {\it decrease} of the $h\to b\overline b$ signal,
dominant in the Minimal SM, and this can adversely affect Higgs searches in $e^+e^-$
colliders. On hadron colliders the increased gluonic coupling enhances Higgs production in
gluon fusion. At the Tevatron, perhaps the most interesting enhancement occurs
in~$gg\rightarrow h\rightarrow\tau^+\tau^-$ for a light Higgs ($m_h<135\,GeV$), while
heavy Higgs production ($m_h>135\,GeV$) will be enhanced in~$gg\rightarrow h\rightarrow
W^+W^-,\ WW^*$.

\section{Mixing at the Lagrangian level}\label{sec:mixing}

The most general form of the relevant part of the Lagrangian after the new interactions have
been integrated out is
\begin{equation}
{\cal L}_{mix}=\frac{1}{2}\partial_\mu\Phi_0^T\cdot
S\cdot\partial^\mu\Phi_0-\frac{1}{2}\Phi_0^T\cdot
M_0^2\cdot\Phi_0 \mbox{ with }
\Phi_0=\left(\begin{array}{cc}h_0\\\sigma_0\end{array}\right),
\end{equation}
and $S$, $M$ representing real symmetric $2\times2$ mixing
matrices. This can always be diagonalized by a field redefinition
\(\Phi=A^T\times\Phi_0\) satisfying \(A\cdot S\cdot A^T=1\),
resulting in
\begin{equation}\label{eq:lag}
{\cal L}_{mix}=
\left[\frac{1}{2}\left(\partial h\right)^2-\frac{m_h^2}{2}h^2\right]
+\left[\frac{1}{2}\left(\partial\sigma\right)^2-\frac{m_\sigma^2}{2}\sigma^2\right]
+h\,j_h+\sigma\,j_\sigma.
\end{equation}

The mixing can be parametrized by four parameters, $\vartheta, \theta,
\rho_h$ and $\rho_\sigma$,
\begin{equation}
A=\left[\begin{array}{cc}
\rho_h\cos\theta&-\rho_\sigma\sin\vartheta\\
\rho_h\sin\theta&\rho_\sigma\cos\vartheta
\end{array}\right].
\end{equation}

Note that the kinetic mixing with these parameters is
\begin{equation}
S=(A^T\cdot A)^{-1}=\frac{1}{\cos^2(\theta-\vartheta)}\left[
\begin{array}{cc}
1/\rho_h^2  & \sin^2(\theta-\vartheta)/\rho_h\rho_\sigma
\\
\sin^2(\theta-\vartheta)/\rho_h\rho_\sigma & 1/\rho_\sigma^2
\end{array}\right],
\end{equation}
so that in the absence of kinetic mixing the parameters reduce to only one
mixing angle through $\theta=\vartheta$ and $\rho_h=\rho_\sigma=1$.

The interactions of the physical Higgs and $\sigma$ are described by the
currents in~(\ref{eq:lag}),
\begin{equation}
j_h=\rho_h\cos\theta\,j_{h_0}-\rho_\sigma\sin\vartheta\,j_{\sigma_0}
\mbox{ and }
j_\sigma=\rho_\sigma\cos\vartheta\,j_{h_0}+\rho_h\sin\theta\,j_{h_0},
\end{equation}
which in the absence of kinetic mixing reduces to
\begin{equation}
j_h=\cos\vartheta\,j_{h_0}-\sin\vartheta\,j_{\sigma_0}
\mbox{ and }
j_\sigma=\cos\vartheta\,j_{h_0}+\sin\vartheta\,j_{h_0}.
\end{equation}

The effects of such mixing will show up both in $h$ production and decay.

In the following we will focus on the case of a light Higgs with mass $100\,GeV\leq
m_h\leq135\,GeV$, when only the $g,g$ and $b\overline b$ Higgs decay channels are important in
the SM. We will comment on the heavy Higgs in~Sec.~\ref{sec:modif}. The original currents are
\begin{eqnarray}
j_{h_0}&=&y_b\,\overline b^Ab_A\\
j_{\sigma_0}&=&\frac{f_g}{2m_t}\times G^a_{\mu\nu}G_a^{\mu\nu}+
\overline b\left[
f_R\frac{1+\gamma^5}{2}+f_L\frac{1-\gamma^5}{2}
\right]b.
\nonumber\end{eqnarray}
These depend, in general, on three new dimensionless couplings $f_R,f_L$ and $f_g$. In the
above the $h_0gg$ coupling has been neglected, because it only arises from loop diagrams.

After diagonalization the  currents that couple to the physical Higgs and $\sigma$ become
\begin{eqnarray}
j_{h}&=&\overline b\left[
\left(y_b\cos\theta-
f_Rsin\vartheta\right)\frac{1+\gamma^5}{2}+\left(y_b\cos\theta-
f_Lsin\vartheta\right)\frac{1-\gamma^5}{2}
\right]b-
\frac{f_g\sin\vartheta}{2m_t}\times G^a_{\mu\nu}G_a^{\mu\nu},\\
j_{\sigma}&=&
\overline b\left[
\left(y_b\sin\theta+
f_Rcos\vartheta\right)\frac{1+\gamma^5}{2}+\left(y_b\sin\theta+
f_Lcos\vartheta\right)\frac{1-\gamma^5}{2}
\right]b+
\frac{f_g\cos\vartheta}{2m_t}\times G^a_{\mu\nu}G_a^{\mu\nu}.
\nonumber\end{eqnarray}
The partial decay widths of the Higgs and the $\sigma$ are found from these couplings at the
corresponding poles. 

We calculated the total widths of both particles, denoted as $\Phi=h,\sigma$,
coupled to a generic current \(j_\Phi=\frac{f_\Phi}{2m_t}G^2+\overline
b\left(f_R^\Phi\frac{1+\gamma^5}{2}+f_L^\Phi\frac{1-\gamma^5}{2}\right)b\)),
and found
\begin{eqnarray}
\Gamma_{\Phi\to gg}&=&
\frac{m_\Phi}{32\pi}\left(\frac{m_\Phi}{m_t}\right)^2\times
\nonumber\left(N_c^2-1\right)\times|f_\Phi|^2
\\\Gamma_{\Phi\to b\overline b}&=&
\frac{m_\Phi}{32\pi}\times N_c\times\left(|f_L^\Phi|^2+|f_R^\Phi|^2\right).
\end{eqnarray}

The Higgs decay ratio is then found to be
\begin{equation}
R_h\equiv\frac{\Gamma_{h\to gg}}{\Gamma_{h\to b\overline b}}=
\frac{N_c^2-1}{N_c}\left(\frac{m_h}{m_t}\right)^2\times\frac{f_g^2\sin^2
\vartheta}{y_b^2\cos^2\theta+\left(f_L^2+f_R^2\right)\sin^2\vartheta-
2y_b(f_L+f_R)\cos\theta\sin\vartheta}.
\end{equation}

We note that this formula involves essentially only two parameters. This is
manifest if the parameter \(f^2\equiv
f_L^2+f_R^2=\frac{32\pi}{N_c\,m_\sigma}\Gamma_{\sigma_0\to b\overline b}\) is used,
together with the angle $\alpha$, defined as $f_L=f\cos\alpha$, 
$f_R=f\sin\alpha$:
\begin{eqnarray}
R_h&=&\frac{8}{3}\,\frac{m_h^2}{m_t^2}\,\frac{f_g^2\sin^2\vartheta}
{(y_b\cos\theta)^2+(f\sin\vartheta)^2-
2(y_b\cos\theta)(f\sin\vartheta)\times\sqrt2\cos(\pi/4-\alpha)}
\nonumber\\&=&\label{eq:BR}
\frac{\frac{m_h^2}{m_\sigma^2}R_\sigma}{1+
\left(\frac{y_b\cos\theta}{f\sin\vartheta}\right)^2-
2\left(\frac{y_b\cos\theta}{f\sin\vartheta}\right)\times\sqrt2\cos(\pi/4-\alpha)},
\end{eqnarray}
where we introduced the notation \(R_\sigma\equiv\frac{\Gamma_{\sigma\to
gg}}{\Gamma_{\sigma\to b\overline b}}\). From Eqn.~(\ref{eq:BR}) we see that
the term containing the parameter $\alpha$ plays only a minor role
(disregarding the exceptional case when a cancellation occurs in the denominator), and this
fact justifies the approximation in the last line.

We can argue that except for two very unlikely cases the unity in the denominator
of Eqn.~(\ref{eq:BR}) dominates. For these exceptions to happen we need
\(\frac{\sin\vartheta}{\cos\theta}\leq\frac{y_b}{f}\equiv\sqrt{
\frac{m_\sigma}{8\times10^4\Gamma_{\sigma\to b\overline b}}}\). This would require
either (i) a very small mixing angle $\theta<10^{-2}$, or (ii) a very small partial width
$\Gamma_{\sigma\to b\overline b}/m_\sigma\leq10^{-4}$. Both of these cases are hardly
conceivable.

Most probably, the mixing will be large enough for the unity in the
denominator of Eqn.~(\ref{eq:BR}) to dominate. In that case, the
observed Higgs branching ratio is {\it independent} of the mixing angles,
\(R_h\approx\frac{m_h^2}{m_\sigma^2}\times R_\sigma\). The physical
reason for this is that both $h\to gg$ and $h\to b\overline b$ are dominated
by the decay through mixing, $h\to\sigma^*\to gg,b\overline b$.

\section{Estimates of $\sigma$ decays in a potential model}\label{sec:Yukawa}

We estimate the properties and decay widths of the $\sigma$ particle in a
potential model with a Yukawa potential. This particular framework is not necessarily
realized in Nature but we remind ourselves that all we want is an existence proof for a
situation where this $\sigma$ might exist. The advantages of such a potential
are: (1) it has two parameters, both of which can be reasonably closely
related to parameters of the underlying theory (a new gauge
coupling $\alpha_{new}$ and a new symmetry breaking scale
$m$), and (2) that the introduction of ${\bf L}\cdot {\bf S}$ coupling allows for a low lying
$2^{\,3}\!P_0$ state, while all other states become heavy or nonexistent.

We use such a potential model to calculate the toponium spectrum, to restrict
the new coupling strength and screening mass, and to find the decay widths of the
$\sigma$. Especially, we use the usual potential model expression that
relates the decay rate $\sigma\to gg$ to the derivative of the radial wave
function at the origin. To find these quantities we need to calculate the wave
function of a particle moving in a Yukawa potential.

\subsection{The Yukawa problem}

Consider the quantum mechanical problem of a point particle with mass $M$
moving in a Yukawa potential with screening mass $m$. The Schr\"odinger
equation for the radial wave function is
\begin{equation}
\frac{u_l''(r)}{u_l(r)}=\frac{l(l+1)}{r^2}-2M\left({\cal E}+\frac{e^{-mr}}{r}\right).
\end{equation}
In our case of a $t\overline t$ system M is related by a rescaled reduced top
mass by \(2M=4\pi\alpha_{new}m_t\). This relationship is at best only
approximate, so that the parameter $\alpha_{new}$ above may not exactly
coincide with the new gauge coupling. The binding energy of the $t\overline
t$ bound state is then $E=4\pi\alpha_{new}{\cal E}$.

The energy levels in the Yukawa potential have been calculated with high precision
in~\cite{Stubbins,Vrscay}. Because in these references the wave functions are not quoted, we
repeated the calculation of Ref.~\cite{Stubbins}, where $u_l(r)$ is approximated by a linear
combination of functions$~r^{k+1}e^{-\frac{\beta}{2}r}$. Then the coefficients of the linear
combination and also $\beta$ are varied in order to minimize the energy of the state.The results are
shown in Table~\ref{table:Yukawa}.
\begin{table}
\begin{tabular}{||c|c|c|c||}
    Screening parameter  &   $m/M=0 $  &   $m/M=0.1$   &   $m/M=0.2$\\  
\hline\hline
$\frac{\cal E}{M}(1S)$   &  -0.4999  &  -0.4071       & -0.3268      \\
$\frac{\cal E}{M}(2S)$   &  -0.1249  &  -0.04993     & -0.1211      \\
$\frac{\cal E}{M}(2P)$   &  -0.1250  &  -0.04653     & -0.004102   \\   \hline
$R_{2P}(0)$                  &   0           &  -0.000,0521&  0.00125     \\
$R_{2P}'(0)$                 &   0.2041  &   0.1802       &  0.1063
\end{tabular}
\caption{\protect\label{table:Yukawa}The energy levels in a Yukawa potential.
The radial wave function $R_l(r)=\frac{u_l(r)}{r}$ and its derivative at $r=0$ is
also given for the $2P$ state.}
\end{table}

As a consistency check, observe that the P-wave wave function at the origin
is found to be extremely small. Because in our approximation this is not
automatic, its smallness indicates that our calculation is highly accurate.
Moreover, we checked that in the $m=0$ case (pure Coulomb potential) we
reproduce the exact results within the accuracy indicated by the valid digits
kept, and also that the approximation to the wave function is better than
$0.2\%$ for all $r$. To the same accuracy, all the energy levels in
Table~\ref{table:Yukawa} co\"\i ncide with the ones given in
Refs.~\cite{Stubbins,Vrscay}.

Inspection of Table~\ref{table:Yukawa} shows that increasing values of the
screening parameter will lessen the binding energies. The critical values at
which each of the states disappears are given in Ref.~\cite{Vrscay}:
\[
\frac{m}{M}(1S)=1.91,\ \ \ 
\frac{m}{M}(2S)=0.31,\ \ \ 
\frac{m}{M}(2P)=0.22.
\]
This will impose an upper bound on the screening mass.

\subsection{Application to $t\overline t$}

We note first of all that the Yukawa potential we are considering will not bind
the $b\overline b$ system, even if the right handed $b$ quarks participated in the new interaction.
The size of such a state in a pure Coulomb potential would be determined by the inverse bottom
mass, the Bohr radius being $a\sim\frac{1}{4\pi\alpha_{new}m_b}$. The screening in the Yukawa
potential ($m_{new}\gg{\cal O}(100\,GeV)$) completely shuts off the interaction
at these distances. Another way to see this is that the rescaled reduced mass
in $b\overline b$ would be $M=4\pi\alpha_{new}\frac{m_b}{2}$. But the new
interaction must be screened much above $m\gg100\,GeV$, and in order for
a bound state to form the screening parameter must satisfy $\frac{m}{M}<1.91$.
These constraints can be satisfied only with $\alpha_{new}\gg2$, an interaction
too strong to be realistic.
\begin{table}
\begin{tabular}{||c|c|c|c|c|c|c||}
Screening parameter&$0 $&$0.10$&$0.16$&$0.18$&$0.20$&$0.22$\\ 
$m/\alpha_{new}$&$0$&$109\,GeV$&$174\,GeV$&$196\,GeV$&$218\,GeV$&$240\,GeV$\\\hline
$E_{bind}\,(\sigma\equiv2P)\,/\,\alpha_{new}^2$&$-1.7\,TeV$&$-640\,GeV$&$-270\,GeV$&$-160\,GeV$&$-56\,GeV$&$0$\\
$E_{bind}\,(1S)\,/\,{\tilde\alpha}_{new}^2$&$-6.9\,TeV$&$-5.6\,TeV$&$-4.9\,TeV$&$-4.7\,TeV$&$-4.5\,TeV$&$-4.3\,TeV$\\\hline
$\left(\frac{\Gamma_{\sigma\to gg}}{m_\sigma}\right)^{1/5}/\left(\frac{m_t}{m_\sigma}
\alpha_{new}\right)$&3.52&3.17&2.58&2.25&1.94&N/A
\end{tabular}
\caption{\protect\label{table:res}The binding energy levels and the $\sigma\to
gg$ width in a Yukawa potential. Recall that the screening parameter,
$\frac{m}{M}\equiv\frac{m}{2\pi\alpha_{new}m_t}$, arises from a new symmetry breaking
scale (related in turn to the mass of the new gauge boson masses.)}
\end{table}

Now we use the Yukawa potential to understand the toponium spectrum as
well as the decay of $\sigma$ into two gluons. The latter is calculated using
the usual QCD formula for quarkonium ($\chi_0\to gg$) decay~\cite{etatogg},
\(\Gamma_{\sigma\to gg}=96\alpha_s^2\frac{|R'(0)|^2}{m_\sigma^4}\).The
results are shown in Table~\ref{table:res}.

From Table~\ref{table:res} we quickly deduce the restrictions on the
parameters when our scenario can occur. Due to the fifth power in the
expression for the $\sigma\to gg$ width, $\alpha_{new}$ is restricted from
above. (Intuitively, the high power of $\alpha_{new}$ arises because large couplings pull together
the wave function closer to the origin and provide a large $|R'(0)|^2$.) Essentially at
$\alpha_{new}>1$, the decay width suddenly becomes much larger than $m_\sigma$, so that the
$\sigma$ resonance is not formed. On the other hand, we must have $m\gg 100\,GeV$ if we want to
avoid a large contribution to top physics, which requires
$\alpha_{new}>0.5$ (See the second row in Table~\ref{table:res}.) 
One conceivably realistic value is
$\alpha_{new}=0.7$ and $m=140\,GeV$, which would result in a binding energy of
$-80\,GeV$, i.e. $m_\sigma=270\,GeV$ and $\Gamma_{\sigma gg}=300\,GeV$.

There is one problem with the above situation. The P-wave state is
necessarily in the second shell (corresponding to a $2^{\,3}\!P_0$), and in the
realistic parameter range the first shell ($1S$) is very tightly bounded. Its
mass is actually negative, which triggers condensation of the vacuum. This is
reminiscent to a topcolor or top condensation scenario~\cite{R7}, which is outside the scope of the
present paper, where we consider electroweak symmetry breaking through the usual
Higgs mechanism.

We should note that this conclusion is brought about by the assumption that
the new interaction does not generates a spin flip. That this assumption is
unrealistic is shown by the simple analogy to nuclear forces, where there is a large
spin-orbit interaction which even changes the ordering of the energy levels.

Fortunately, the inclusion of a simple ${\bf L}\cdot {\bf S}$ interaction quickly remedies
the situation. Assume that the Yukawa potential includes such a term (we use
the same screening mass in all terms for simplicity)
\begin{equation}
V(r)=-g^2\frac{e^{-mr}}{r}\left(\cos\gamma-{\bf L}\cdot{\bf S}\sin\gamma\right).
\end{equation}
Then all the  $n^{\ 2S+1}\!L_J$ states are ${\bf L}\cdot{\bf S}$ eigenstates, so
that they feel a pure Yukawa potential (but for each state the strength of the
coupling it feels, $\alpha_{eff}$, is different, as shown in Table~\ref{table:strength}).
\begin{table}[b]
\begin{tabular}{||c|c|c|c|c|c|c||}
State&$^1S_0$&$^3S_1$&$^1P_1$&$^3P_2$&$^3P_1$&$^3P_0$\\
${\bf L}\cdot{\bf S}$&0&0&0&+1&-1&-2\\
$\alpha_{eff}/\alpha_{new}$&
   $\cos\gamma$&$\cos\gamma$&$\cos\gamma$&
   $\cos\gamma-\sin\gamma$&$\cos\gamma+\sin\gamma$&
   $\cos\gamma+2\sin\gamma$
\end{tabular}
\caption{\protect\label{table:strength}The binding energy levels and the
$\sigma\to b\overline{b}$ width in a Yukawa potential.}
\end{table}
Because the state that is most sensitive to the spin-orbit interaction
happens to be the $\sigma$ (i.e. $^{\,3}\!P_0$), we can shift all other states
to smaller binding energies as compared to~$\sigma$. Replacing the
parameter $\alpha_{new}\to\alpha_{eff}$, the numbers in Table~\ref{table:res} remain in
force. The first shell with only S-wave states feels a different, weaker potential with
$\alpha_{eff}(1S)=\frac{\alpha_{new}}{1+2\tan\gamma}$. If the potential is
dominated by the spin-orbit interaction, the binding energy of the $1S$ states
can be made small. Because the screening parameter is also affected by the
new $\alpha_{eff}$ (while the screening mass $m=140\,GeV$
remains unchanged), in the numerical example given above, for
$-1<\cos\gamma<0.2$, the {\it only} remaining state is the
$\sigma\equiv2^{\,3}\!P_0$.

\section{Modifications to the Higgs signals}\protect\label{sec:modif}

We have shown an example when the mixing of the Standard Model Higgs boson with a
new heavy scalar introduces significant changes in the couplings of the Higgs,
most notably leading to a sizable, perhaps dominant, Higgs-gluon interaction. In
the  Higgs mass region $115\,GeV<m_h<180\,GeV$ that we are considering, this changes the
relative importance of the various production channels as well as the decay modes. This alters the
strategy needed to discover the Higgs boson in a significant way, both at the hadron and lepton
colliders.

In future leptonic colliders the environment is favorably clean for the
detection of the Higgs. The ``gold plated" mode $e^+e^-\rightarrow Z^*\rightarrow Zh$
followed by the Higgs decay $h\rightarrow b\overline{b}$ seems to be the easiest to see. The
mixing discussed in this paper would increase the $Z$ plus two gluon jets final state. A
large $BR(h\rightarrow gg)\sim 1$ branching ratio arises if the new scalar $\sigma$
dominantly decays into two gluons. This would diminish the $Zb\overline b$ signal so that
the Higgs might be missed altogether in searches for a standard Higgs. This is also important for the
LEP2 data with the search restricted to $h\to$ two $b$-jets.

At Tevatron Run~2, the most promising mode for the SM Higgs boson in the mass range
$110\,GeV\leq m_h\leq130\,GeV$ is $q\overline q^\prime\to W^*\to Wh\to Wb\overline b$.
However, in our scenario, the dominant decay mode of $h$ may be $h\to gg$, thus reducing the
$h\to b\overline b$ branching ratio. So the signal in the $Wb\overline b$ mode may be greatly
reduced, making it unobservable due to the large QCD background. On the other hand, the Higgs
production in the mode $gg\to h$ will be greatly enhanced in this scenario. Thus, the signal in the
$gg\to h\to WW^*$ channel will now be greatly enhanced. This will extend the lower mass range
coverage below the usual case of $m_h\sim135\,GeV$ via this mode. The mode $gg\to
h\to\tau^+\tau^-$ was a borderline case in the SM. With the enhanced $gg\to h$ production rate of
our scenario, the signal in this mode may be observable if the $\tau$ detection efficiency and the
missing $E_T$ resolution can be improved in the upgraded CDF and D0 detectors.

In the SM, $\sigma(gg\to H)\sim0.7\,pb$, while $B(h\to\gamma\gamma)\sim2.2\times10^{-3}$ for
$m_H=120\,GeV$. In our scenario, the Higgs branching ratio to two photons will be increased, and so
will be the $\sigma(gg\to H)$ production cross section. Thus, the two photon mode may be
detectable at Tevatron Run~2. A similar enhancement will occur for the two photon mode at the
LHC.

While in most of this paper we discussed the case of a light Higgs boson, it is worthwhile to
point out that  in the case of a heavy Higgs ($m_h\geq135\,GeV$, when the dominant decay
is $h\rightarrow WW,WW^*$) the QCD background is less severe on hadronic colliders. Then
gluon fusion is the dominant process in the Standard Model, $gg\rightarrow h\rightarrow
W^+W^-,\ WW^*$, providing an excellent signature. This signal will be greatly enhanced in the
proposed scenario.

\section{Conclusions}

We have presented a scenario in which the coupling of the SM Higgs boson to two gluons can be
greatly enhanced due to mixing with a scalar $t\overline t$ bound state (bound by new strong
interactions). We have presented a potential model in which such a scenario may be realized. Such a
scenario has major implications for the Higgs search at Tevatron Run~2 in the mass range of
$110\,GeV\leq m_h\leq180\,GeV$. Production and decay channels are both affected. The
$Wb\overline b$ final state at Tevatron Run~2 will be suppressed, because the $h\to b\overline b$
branching ratio is suppressed. The $WW^*$ mode will be enhanced due to an increase in the $gg\to
h$ production cross section, so that this mode will be effective even for a Higgs mass range lower
than $135\,GeV$. Because Higgs production via $gg\to h$ mode is enhanced, the $\tau^+\tau^-$
and perhaps also the $\gamma\gamma$ final states may be detectable. Finally, we point out that
LEP2 may have missed the Higgs boson because the dominant final states were $Zgg$ instead of
$Zb\overline b$.

After the completion of this paper another work was published~\cite{HewettRizzo}, which discusses
a similar effect on the gluonic couplings of the Higgs boson. The underlying model in that paper is
based on extra dimensions and is very different from our scenario.

\section*{Acknowledgments}

This work has been supported by the US Department of Energy Grant Nos. DE-FG03-98ER41076 and
DE-FG02-01ER45684.

\end{document}